# Absolute Radar-REMPI via Two-Color REMPI and Absolutely Calibrated CMS for Diagnostics of Species in Gaseous Mixtures

Mathis Malaussena[1], Liam West[2], Kenneth Reindersma[3], Nicholas Babusis[4], Alexey Shashurin[5]
*Purdue University, West Lafayette, IN 47907, USA*

**This work demonstrates the initial development of the Absolute Radar-REMPI diagnostic technique, a development of the Radar-REMPI technique that provides an approach for absolutely calibrated measurements, which is universally applicable to any type of tested gaseous species. Absolute measurements are achieved through the utilization of two-color Resonance Enhanced Multiphoton Ionization (REMPI) and absolutely calibrated Coherent Microwave Scattering (CMS). Two-color REMPI of molecular oxygen ($O_2$) was demonstrated using two intersected laser beams: a 355 nm beam and a tunable parametric beam at 241-242 nm. The results confirm that the two-beam configuration can successfully generate measurable ionization associated with the targeted transition of $O_2$ within the beam intersection region, as verified using an ICCD camera. These findings establish a foundation for the further development of Absolute Radar-REMPI, which has the potential to enable highly sensitive, real-time diagnostics of absolute species densities in unsteady flowfields such as those occurring in hypersonic shock tunnels and electric propulsion systems.**

## I. Nomenclature

| | | |
|---|---|---|
| *CMS* | = | Coherent Microwave Scattering |
| *REMPI* | = | Resonance-Enhanced Multiphoton Ionization |
| $N_e$ | = | Total Number of Electrons |
| $n_s$ | = | Number Density of Species |
| $V$ | = | Plasma Volume |
| *CCD* | = | Charge-Coupled Device |
| *ICCD* | = | Intensified Charge-Coupled Device |

[1] MS Student, Purdue University, School of Aeronautics and Astronautics.
[2] BS Student, Purdue University, School of Aeronautics and Astronautics.
[3] MS Student, Purdue University, School of Aeronautics and Astronautics.
[4] PHD Student, Purdue University, School of Aeronautics and Astronautics.
[5] Associate Professor, Purdue University, School of Aeronautics and Astronautics.

## II. Introduction

Radar-REMPI (Resonance Enhanced Multiphoton Ionization) is an effective diagnostic technique for measurements of selective species in gaseous mixtures.[1,2] However, the conventional Radar-REMPI technique is inherently a relative measurement technique and cannot provide absolutely calibrated measurements in the case when the target species do not exist in stable form under normal conditions (such as atomic oxygen, unstable molecules, ionized species, etc.) and absolute calibration is not feasible.

In this paper we propose a novel Absolute Radar-REMPI technique, that offers an approach for absolutely calibrated measurements which is universally applicable to any tested gaseous species. The approach relies on utilization of two-color REMPI and absolutely calibrated Coherent Microwave Scattering (CMS). Here, we demonstrate the first step toward establishing the technique, namely the feasibility of confining the ionization volume to a fixed beam-intersection region using two-color REMPI scheme of oxygen at 355nm and 241-242nm.

## III. Methodology

A schematic of the experimental system is shown in Figure 1. The system consisted of a nanosecond laser, an ICCD camera, and a CMS technique used to probe the laser-induced plasma. The laser source was a modified Ekspla NT 342 pulsed nanosecond laser capable of simultaneously generating two beams of different wavelengths: a fixed 355 nm beam and a tunable optical parametric beam at a variable wavelength in the range of 210-2600 nm. The two beams were intersected at their focal points, where REMPI was generated. The ICCD camera (Pi-Max4 1024i) was positioned directly above the beam intersection region with its optical axis normal to the plane defined by the laser beams. The camera was synchronized with the laser system and operated using background subtraction to automatically remove ambient light contributions from the recorded images. The minimum gate width of 3 ns, combined with adjustable trigger delay, enabled images to be taken at different times during or after the laser pulse. The detection of the laser-induced plasma was performed using a CMS system. The CMS system utilized a homodyne-type detection circuit and a microwave source operating at a frequency of 11 GHz. The transmission (Tx) and receiving (Rx) microwave horns were positioned at a 90˚ angle relative to each other facing the laser-induced plasma at the beam intersection. A microwave absorbing panels were installed around the measurement region to minimize reflections and improve signal quality.

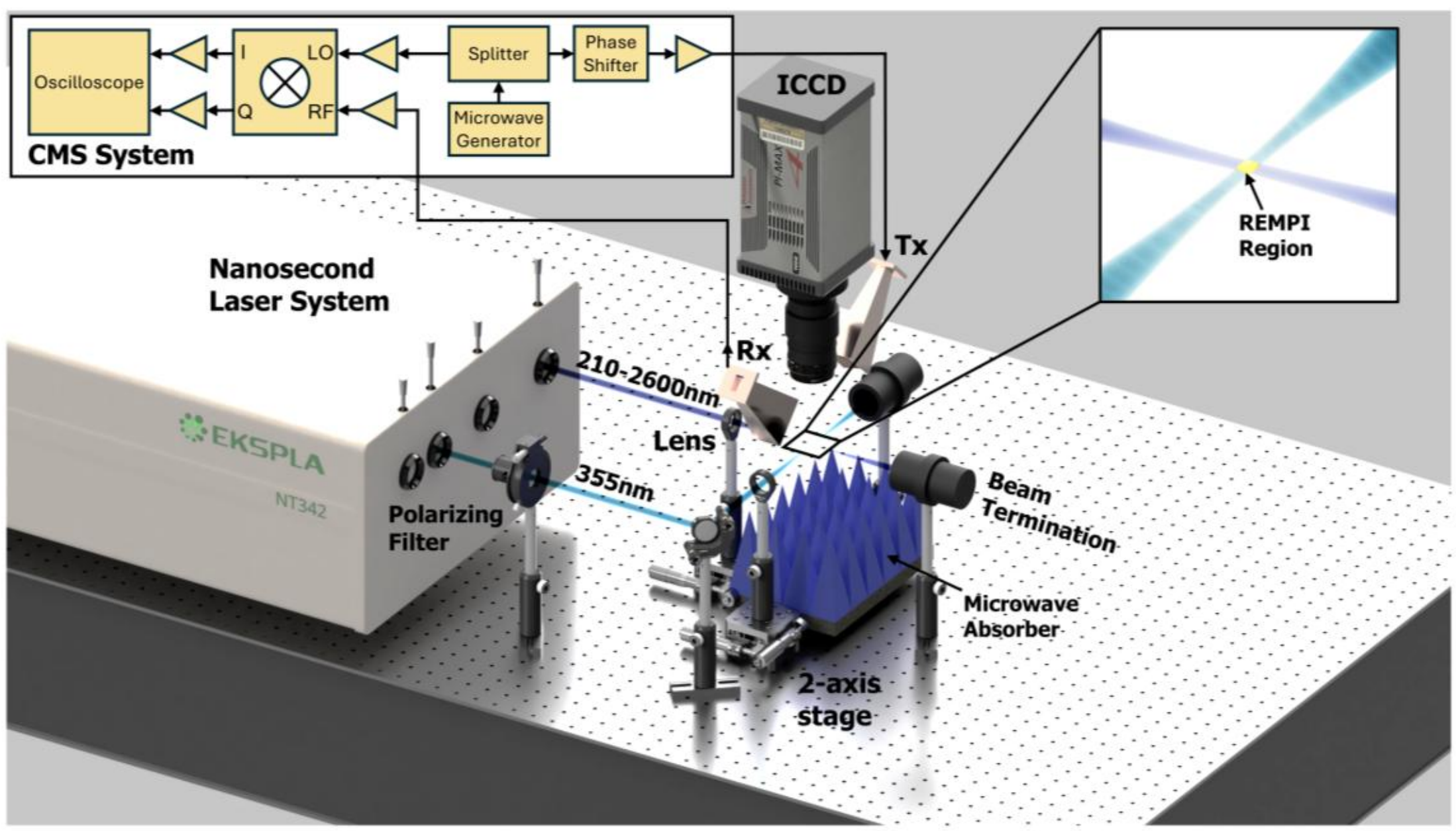

**Figure 1. Experimental implementation of Absolute Radar-REMPI through the integration of two-color REMPI with the absolutely calibrated CMS technique.**

## IV. Results

The proposed diagnostic represents an absolutely calibrated extension of the previously demonstrated Radar-REMPI technique,[1,2] achieved through the utilization of absolutely calibrated CMS and two-color REMPI (accordingly, the technique is referred to as Absolute Radar-REMPI or Absolute CMS Two-Color Radar-REMPI). The absolute calibration capability is particularly important for the measurement of unstable or transient species that do not exist in a stable form under ambient conditions (e.g., atomic oxygen, short-lived molecular species, and ionized species).

The idea of the approach proposed here is to achieve full ionization of a tested species within a fixed volume ($V$) using REMPI, measure the total electron count in that volume ($N_e$) using absolutely calibrated CMS technique, and subsequently calculate the local number density of the tested species using the simple expression $n_s = \frac{N_e}{V}$. To achieve full ionization of the tested species within a fixed volume, we employ two-color REMPI scheme, in which each beam is individually non-resonant, such that REMPI conditions are established only within the fixed intersection region of the two beams. It is important to note that a conventional, single-color REMPI approach would not be suitable here because the volume containing fully ionized target species would not remain fixed as the laser intensity is varied. Instead, this region would continuously extend with laser intensity along the direction of laser beam propagation, resulting in an unknown volume containing fully ionized target species (consequently, the term $V$ in the denominator of the equation above would be ill-defined, preventing accurate evaluation of the local species number density).

In this work we achieve two-color REMPI of oxygen using two laser beams operating around 355 nm and 241.5 nm intersected as shown in Figure 1. From the energy conservation standpoint, simultaneous absorption of 355 nm and 241.5 nm photons is equivalent to excitation by two 287.5 nm photons: $\mathrm{O_2}\ \mathrm{C^3\Pi_g}\ (\mathrm{v'} = 2,\ \mathrm{J'}) \leftarrow \mathrm{O_2}\ \mathrm{X^3\Sigma_g^-}\ (\mathrm{v''} = 0,\ \mathrm{J''})$. Experimentally, the two beams were passed through 100 mm lenses at a 90° angle relative to each other and intersected at their focal points as shown in Figure 1 (a polarizing filter was placed on the 355 nm beam in order to reduce its intensity and ensure that it's not producing significant ionization on its own). Spatial alignment of the beams was verified using the CCD sensor of an Allied Vision ALVIUM camera placed directly at the beam-intersection region. The two lasers were dimmed, and their focal points were then aligned on the CCD sensor using a two-axis translation stages for the lenses in combination with a kinematic mirror mount. Following this initial alignment, the ICCD camera was employed to perform fine spatial and temporal alignment of the beams.

The ICCD images of the afterglow plasma generated by the laser pulses used in this work are presented in Figure 2. Specifically, the top row shows the image of plasma generated by the 355 nm beam (8.6 mJ per pulse), the middle row shows plasma generated by the 241–242 nm beam (3.0 mJ per pulse), and the bottom row shows plasma generated when both beams were applied simultaneously. It can be seen that the highest plasma emission intensity (and, consequently, the highest plasma density) was observed at the beam intersection region. To confirm the occurrence of the two-color REMPI process, beyond simple linear superposition of the non-resonant photoionization produced by the individual beams, we analyzed the average ICCD pixel brightness within the beam intersection region, as shown in Figure 3. One can see that ICCD pixel brightness associated with the plasma in the beam intersection region was significantly greater than the linear sum of the signals produced by the individual beams. This observation supports the occurrence of the two-color REMPI process within the beam intersection region. Finally, the presence of the substantial two-color REMPI throughout the 241–242 nm wavelength range may be attributed to the relatively broad resonance line expected for this process, which was previously reported to have a linewidth of approximately 1.3 nm.[3]

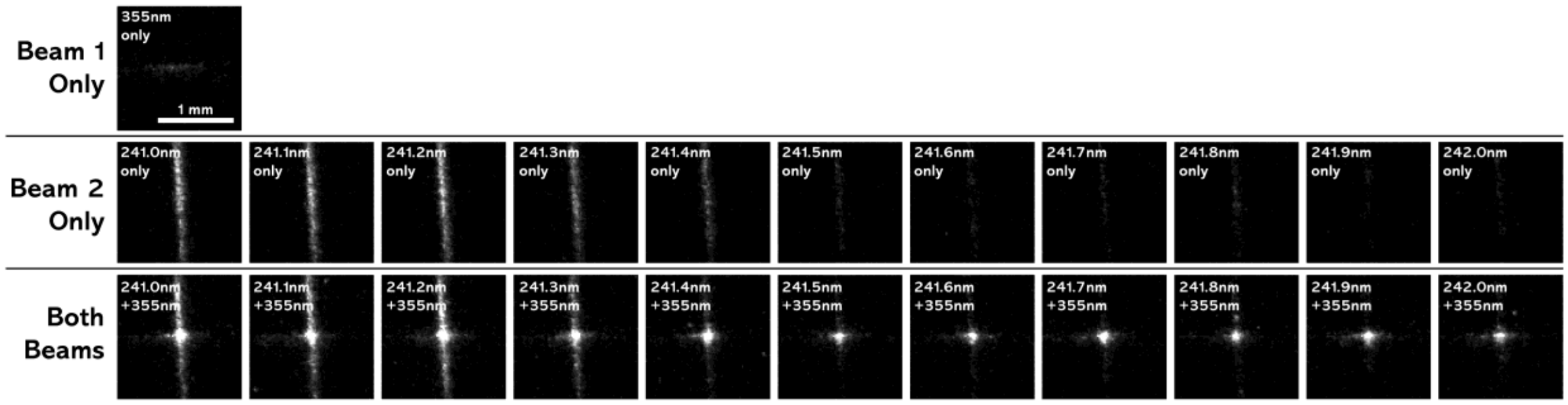

**Figure 2. ICCD images of the afterglow plasma generated by the 8.6 mJ pulse at 241-242nm, by the 3.0 mJ pulse at 355 nm, and by both beam simultaneously (two-color REMPI).**

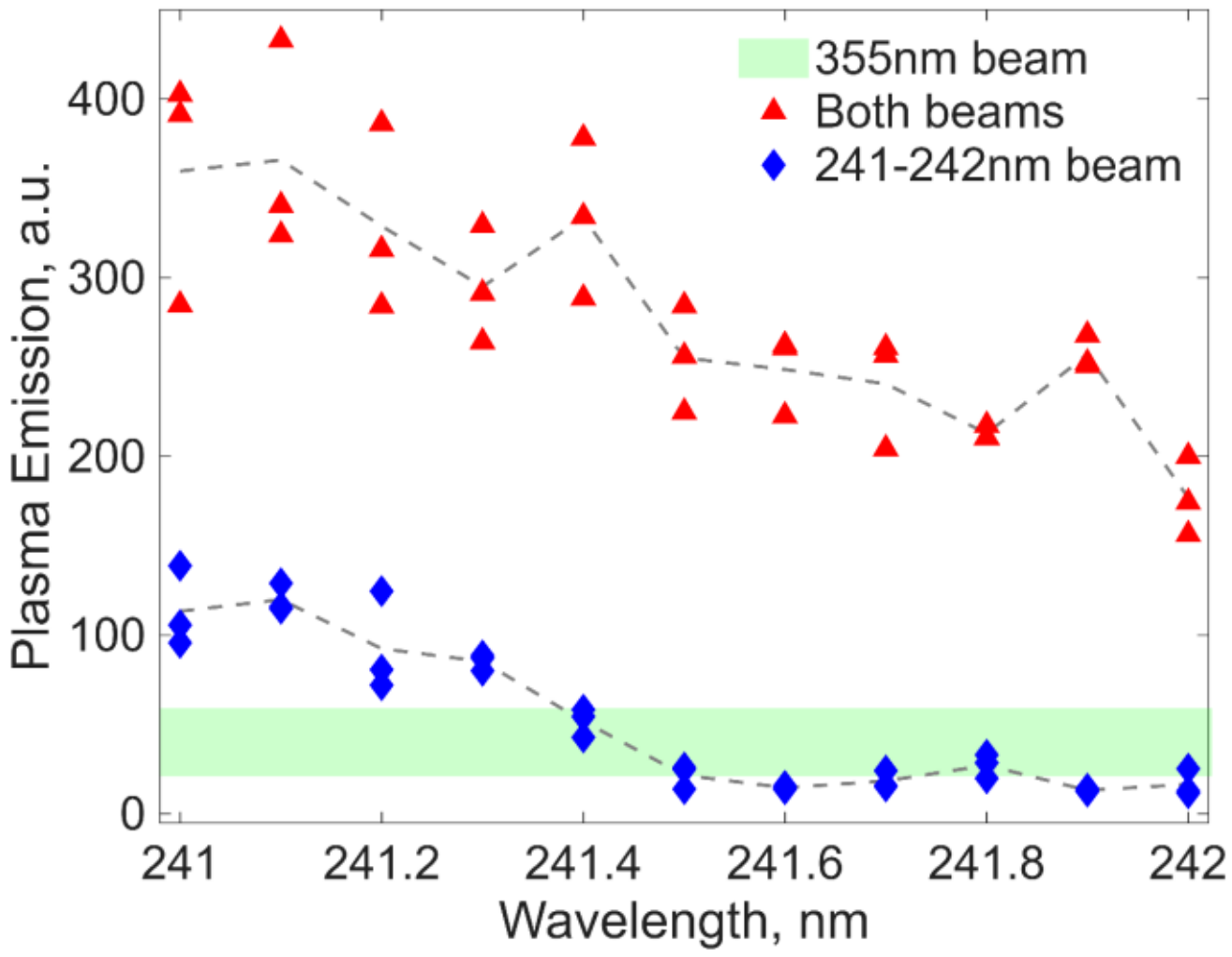

**Figure 3. Emission of the laser-induced plasma generated by the 8.6 mJ pulse at 355 nm, by the 3.0 mJ pulse at 241-242 nm, and by both beam simultaneously (two-color REMPI).**

It is important to note that, with current laser system (Ekspla NT 342), achieving the required full ionization of target species would not be feasible, and two modifications of the current experiment are needed. First, the REMPI process utilized in this study is not sufficiently efficient, as evidenced by the relatively dim plasma emission observed in the ICCD images even at the highest laser pulse energy, indicating a relatively low degree of ionization. Therefore, a different REMPI scheme with a substantially higher photoionization rate is required to be used. One promising candidate is the (2+1) REMPI of krypton at 212.5 nm, which has been observed to produce a significantly brighter plasma volume than the (2+1) REMPI of oxygen at 287.5 nm using similar laser intensities. A typical CMS signal obtained from single-color (2+1) REMPI of krypton at 212.5 nm is shown in Figure 4. One can see a very sharp REMPI resonance with a linewidth of less than 0.1 nm around 212.5 nm. Second, even when employing a highly efficient REMPI process (associated with high photoionization rate), experiments must be conducted at reduced pressures because nonlinear optical phenomena significantly limit the achievable degree of ionization when atmospheric pressure is used. For example, photoionization of atmospheric pressure air by 800 nm photons has been reported to be limited to ionization fractions below approximately 1% due to nonlinear optical effects.[4] At the same time, reduced pressure delays the onset of these nonlinear effects to higher intensities and, correspondingly, enables a higher degree of ionization.[5] This refers to the fact that nonlinear optical phenomena including Kerr and plasma nonlinearities reduce proportionally with a decrease in pressure.

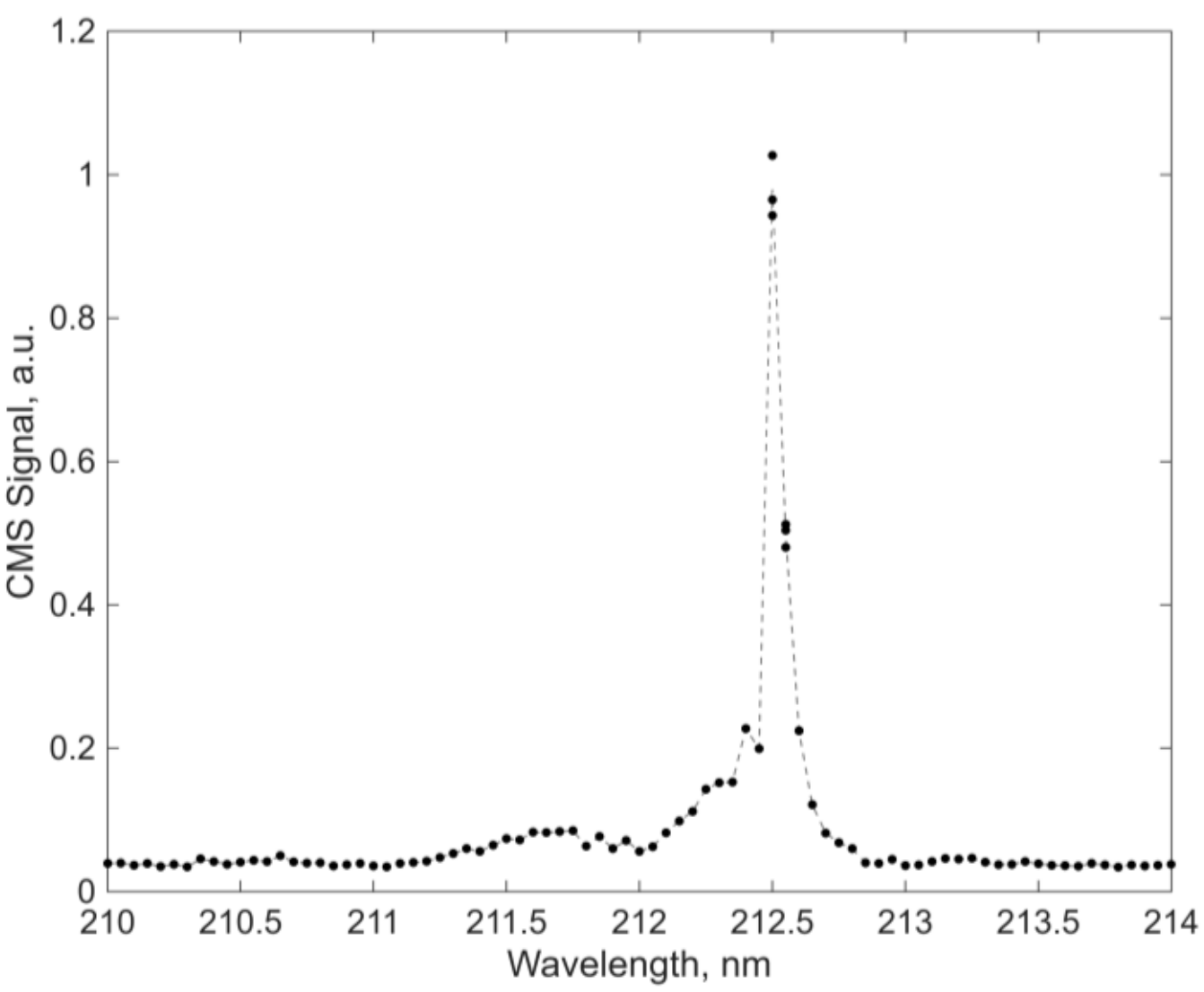

**Figure 4. A typical CMS signal obtained from single-color (2+1) REMPI of krypton at 212.5 nm.**

Therefore, future experiments will employ a more efficient REMPI scheme and will be performed at reduced pressures. Specifically, the two-color REMPI of krypton by 212.8 nm (the fifth harmonic of the 1064 nm fundamental) and 212.2 nm photons will be studied. This excitation scheme is energetically equivalent to the conventional single-color (2+1) REMPI of krypton at 212.5 nm. Finally, the studies will be performed in the vacuum facility shown in Figure 5. The facility consists of a 24-inch cubic stainless steel vacuum chamber whose walls are covered with microwave absorber panels to provide an anechoic environment suitable for isolating the scattered microwave signal. The chamber is evacuated using a diffusion pump and is equipped with fused-silica viewports for laser access, microwave instrumentation feedthroughs, and additional ports for gas handling and auxiliary instrumentation. The chamber can be filled with a variety of gases at controlled pressure levels.

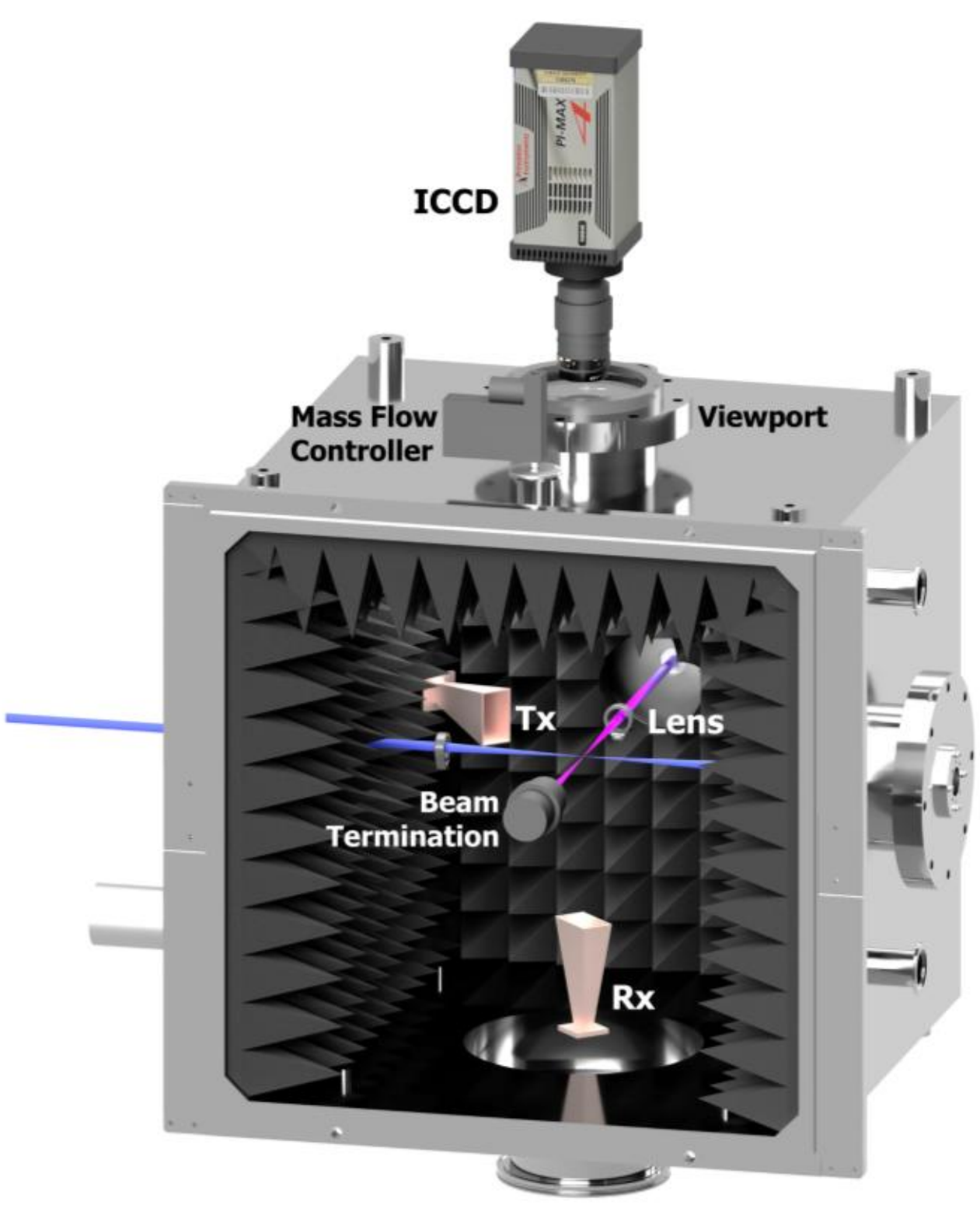

**Figure 5. Experimental facility planned for future development of Abslute Radar-REMPI, enabling testing in a variety of pure-gas environments and identification of efficient REMPI schemes.**

Let us now discuss the expected performance of the proposed Absolute Radar-REMPI diagnostics. The key features of the technique include the utilization of an absolute calibration approach applicable to any type of tested species, real-time measurements (without signal accumulations), high sensitivity, nanosecond-scale temporal resolution and millimeter-scale spatial resolution. Specifically, the proposed absolute calibration approach is universal and can be applied to any type of tested species, provided that an appropriate REMPI scheme is identified for the species of interest. The minimal measurable species number density should be expected to be on the order of $10^{10}$-$10^{11}$ $cm^{-3}$, assuming standard level of the CMS technique sensitivity of about $10^8$ electron and a probing volume of about 1 $mm^3$.[6] The temporal resolution of the method is limited by the slower of two timescales: the temporal resolution of the CMS techniques, which is typically in sub-nanosecond range,[6] and the laser pulse duration. The spatial resolution is governed by the size of the REMPI region and is expected to be in the millimeter-to-submillimeter range. Importantly, these measurements can be acquired in real time without the need for signal accumulation and averaging.

These anticipated capabilities of Absolute Radar-REMPI enable a wide range of potential applications in which real-time diagnostics of absolute species densities in unsteady gaseous mixtures are sought. One example is the diagnostics of hypersonic flows, where single-shot measurements are essential because many facilities, such as hypersonic shock tunnels, inherently operate in a single-shot mode [7]. Another important application is diagnostics of plumes of electric propulsion systems, where real-time, temporally resolved measurements are required to accurately characterize unsteady plasma dynamics and instabilities.

## V. Conclusion

This work demonstrates the initial implementation of the Absolute Radar-REMPI diagnostic technique, a development of the Radar-REMPI technique that provides an approach for absolutely calibrated measurements, which is universally applicable to any tested gaseous species. Selective ionization of molecular oxygen in air was successfully generated using two-color REMPI by intersecting laser beams at 355 nm and 241-242 nm verified through ICCD imaging. The results confirm the feasibility of confining ionization only at the intersection of the two beams and establish a foundation for the future experiments achieving the full ionization of krypton at reduced pressures. Absolute Radar-REMPI technique has the potential to enable highly sensitive, real-time diagnostics of absolute species densities (down to $10^{10}$-$10^{11}$ $cm^{-3}$ with nanosecond-scale temporal/millimeter-scale spatial resolution) in unsteady flowfields such as those occurring in hypersonic shock tunnels and electric propulsion systems.

## Acknowledgments

This work was supported by the National Science Foundation (Grant No. 2409559).